\documentclass[amsmath,amssymb,eps,12pt]{article}
\usepackage{amsfonts,amssymb,amsmath,epsfig,epic,color}
\usepackage{graphicx}
\usepackage{graphicx}
\usepackage{dcolumn}
\usepackage{bm}

\newcommand{\be}{\begin{equation}}
\newcommand{\ee}{\end{equation}}
\newcommand{\bea}{\begin{eqnarray}}
\newcommand{\eea}{\end{eqnarray}}

\newcommand{\hbarm} { \frac{\hbar^2}{2m} }

\newcommand{\hbarmr}
{ \frac{\hbar^2}{2 m R^2} }

\begin{document}

\title{Finitely Many Dirac-Delta Interactions on Riemannian Manifolds}

\author{\centerline {\small Bar\i\c{s} Altunkaynak$^{1}$, Fatih Erman$^2$, O. Teoman Turgut$^{2,\,3}$}
\\\and
{\scriptsize{$^1$Department of Physics, Northeastern University,
Boston, MA 02115, USA.}} \and {\scriptsize{$^2$ Department of
Physics, Bo\u{g}azi\c{c}i University, Bebek, 34342, \.Istanbul,
Turkey}}
\\\and
{\scriptsize{$^3$Feza G\"{u}rsey Institute, Kandilli, 81220
\.Istanbul, Turkey}}
\\
{\scriptsize{E-mail: altunkaynak.i@neu.edu,
fatih.erman@boun.edu.tr, turgutte@boun.edu.tr}}}

\date{\scriptsize{\textsc{\today}}}

\maketitle


Pacs Numbers: 11.10.Gh, 03.65.-w, 03.65.Ge

\begin{abstract}
This work is intended as an attempt to study the non-perturbative
renormalization of bound state problem of finitely many
Dirac-delta interactions on Riemannian manifolds, $\mathbb{S}^2$,
$\mathbb{H}^2$, and $\mathbb{H}^3$. We formulate the problem in
terms of a finite dimensional matrix, called the characteristic
matrix $\Phi$. The bound state energies can be found from the
characteristic  equation $ \Phi(-\nu^2) A = 0$. The characteristic matrix can be found after a regularization
and renormalization by using a sharp cut-off in the eigenvalue spectrum of the Laplacian, as it is done in the
flat space,  or using the  heat kernel method. These two approaches are equivalent in the
case of compact manifolds. The heat kernel method has a general advantage to find lower bounds on the spectrum
even for compact manifolds as shown in the case of  $\mathbb{S}^2$.
The heat kernels for $\mathbb{H}^2$, and
$\mathbb{H}^3$ are known explicitly, thus we can calculate  the characteristic matrix $\Phi$.
Using the result, we give lower bound estimates of the discrete spectrum.
\end{abstract}

\maketitle

\section{Introduction \label{introduction}}

It is well-known that the exactly solvable Dirac-delta
interactions on the plane and 3-dimensional Euclidean space in
quantum mechanics give rise to some unphysical results for
physical observables, i.e., bound state energy and scattering
cross section are infinite and the problem is said to be
ultraviolet divergent. Nevertheless, there is a systematic way to
dispense with these infinities by means of a so-called
regularization and renormalization, which is first introduced in
quantum field theory for the same reason. This problem constitutes
an analytical example of regularization and renormalization in
quantum mechanics so that it helps us to understand and deal with
it in a more elementary context rather than field theory and it
has been studied in the literature from several point of views
\cite{BFT} - \cite{Nyeo}. Moreover, a single point interaction in
two dimensional flat space is an instructive example of
dimensional transmutation in non-relativistic quantum mechanics
\cite{Thorn, Coleman, Huang, Camblong}. That is, the original
Hamiltonian does not contain any intrinsic energy scale due to the
dimensionless coupling constant in natural units. Nevertheless, a
new  parameter $\mu^2$, which species the bound state energy, must
be introduced after the renormalization procedure which then fixes
the energy scale of the system. (A detailed discussion of
dimensional transmutation in nonrelativistic quantum mechanics is
given in a relatively recent article \cite{Camblong}).

In this study, we consider a bound state problem in which a
non-relativistic particle living in a Riemannian manifold (in
particular $\mathbb{S}^2$, $\mathbb{H}^2$, and $\mathbb{H}^3$)
interacts with finitely many Dirac-delta interactions. Similar to
the corresponding bound state problem on $\mathbb{R}^2$ and
$\mathbb{R}^3$, we encounter divergences in this case as well. The
main purpose of this paper is to show how to non-perturbatively
regularize and renormalize the problem by means of heat kernel
(even in the case where we do not have an explicit expression for
it). After the renormalization, we estimate a lower bound for the
ground state energy for each particular Riemannian manifold. This
problem on two dimensional Riemannian manifolds, such as
$\mathbb{S}^2$ and $\mathbb{H}^2$ also displays a kind of
dimensional transmutation \cite{Camblong}, where new energy scales
different from the intrinsic energy scales of the system appear
after the renormalization. We will briefly discuss it in sections
\ref{heatkernel for S^2} and \ref{delta on H^2}.

Many body version of this problem on $\mathbb{R}^2$ and $\mathbb{R}^3$ is known as the formal non-relativistic limit of
the $\lambda \phi^4$ scalar field theory in (2+1) and (3+1) dimensions. All these are extensively discussed in
\cite{Rajeev2}. Our primary motivation here is coming from the question how the renormalization method for the singular
interactions in quantum mechanics would be performed on Riemannian manifolds, hoping that this may help us to
understand the problem in the realm of quantum field theory. However, we shall postpone the discussion of the many body
extension of it for future work and study first the one-particle Schr\"{o}dinger problem.

The paper is organized as follows. In section \ref{rencompact}, we first define the bound state problem on compact and
connected Riemannian manifolds and reformulate the problem in terms of a finite dimensional matrix $\Phi$, which we
will call as the characteristic matrix \cite{Rajeev2}. Then, we emphasize the relation of the characteristic matrix
with the corresponding spectral functions, resolvent and heat kernel. This allows us to reformulate the renormalization
in terms of heat kernel. After that we continue to the discussion in the following sections by working out concrete
examples. In section \ref{Delta Interactions on Sphere $S^2$}, we consider the delta interaction problem on
$\mathbb{S}^2$ as an example for compact and connected manifolds. Considering the properties of the operator $\Phi$ and
using some properties and upper bound estimates of the heat kernel, Ger\v{s}gorin theorem allows us to estimate a lower
bound for the ground state energy of the system. In section \ref{delta on hyperbolic}, we apply the similar
methodology, developed in the section of heat kernel method for $\mathbb{S}^2$, to the non-compact manifolds, such as
$\mathbb{H}^2$ and $\mathbb{H}^3$ and show that the methods developed for compact manifolds work for some particular
non-compact manifolds as well. Therefore, we renormalize the problem on hyperbolic spaces and give estimates on the
ground state energy of each system.

\section{\label{rencompact}
Renormalization of Finitely Many Dirac-Delta Interactions on
Compact and Connected Riemannian Manifolds $(M,g)$}

The canonical quantization on non-trivial manifolds is known to
have some ambiguities in quantum mechanics. For the path integral
approach to the quantum system, the ambiguity in the canonical
formalism is replaced by the undetermined parameter $\lambda$ and
it can take various possible values \cite{de witt}. We remove this
term for simplicity in all our examples, in which the curvature
term is constant and it corresponds to an overall shift in energy
levels so that we can safely set $\lambda$ to be zero.

Now, we consider a non-relativistic point particle living on a Riemannian manifold $M$ interacting with a finite number
of delta interactions located on the manifold and study bound states of the problem. We first investigate the delta
interactions on a compact and connected Riemannian manifold $(M,g)$ without boundary, of dimension $D=2,3$ with the
Riemannian metric $g$. The kinetic energy operator on Riemannian manifold $(M,g)$ is just the Laplace-Beltrami operator
or simply Laplacian, which is defined, in local coordinates $x\equiv(x^1,...,x^D)$ for a neighborhood in the manifold,
as follows:
\be \triangle_{g} = - \frac{1}{\sqrt{\mathrm{det}\,g}}
\sum_{\alpha,\beta=1}^D \frac{\partial}{\partial x^\alpha}
\left(g^{\alpha\beta} \, \sqrt{\mathrm{det} \,g} \;
\frac{\partial}{\partial x^\beta}\right), \ee
where $g_{\alpha\beta}$ is the metric tensor and
$g=(g_{\alpha\beta})$. We shall usually denote the Laplacian as
$\triangle_g$ to specify which metric structure on Riemannian
manifold it is associated with.

The spectral theorem \cite{Rosenberg,Davies} states that the eigenvalue problem $\triangle_g \phi_{l} = \lambda_{l}
\phi_{l}$ on a compact and connected Riemannian manifold $(M,g)$ has a complete orthonormal system of $C^\infty$
eigenfunctions $\phi_0, \phi_1, \dots$ in $L^2(M)$ and the spectrum
$\mathrm{Spec}\,(\triangle_g)\equiv\mathrm{Spec}\,(M,g)=\{\lambda_l\} = \{0 = \lambda_0 < \lambda_1 \leq \lambda_2 \leq
\dots\}$, with $\lambda_l$ tending to infinity as $l \rightarrow \infty$. As a corollary of this theorem, the Laplacian
on $(M,g)$ provides us with all the tools of Fourier analysis, so that we can expand any ``sufficiently good'' function
$\psi(x)$ on $M$ in terms of the complete orthonormal eigenfunctions $\phi_l(x)$
\be \label{expansion1} \psi(x) = \sum_{l \geq 0} C_l \,
\phi_l(x)\;, \ee
with the normalization
\be \nonumber \int_{M}\phi_l(x)\phi_{l'}^{*}(x)\; \sqrt{\mathrm{det}\,g}\;dx^1 \wedge ... \wedge dx^D = \delta_{ll'}\;,
\ee
where $C_l$'s are expansion coefficients. Note that extra labels
in the eigenfunction expansion must be taken into account if the
problem admits degeneracy. Delta functions on $M$ can also assumed
to be represented by these eigenfunctions
\be \label{expansion2} \delta^D(x-a_i) = \sum_{l \geq 0} \phi_l(x)
\phi_l^*(a_i), \ee
with $a_i \in M$ and $\delta^D(x-a_i)$ being the $D$ - dimensional
normalized delta function at point $a_i$,
\be \nonumber \int_{M} \delta^D(x-a_i)
\;\sqrt{\mathrm{det}\,g}\;dx^1 \wedge ... \wedge dx^D =1 \;. \ee
A typical Hamiltonian operator in quantum theory consists of a
kinetic term, the Laplacian $\triangle_g$ with the factor $\hbar^2
/2m$, and a potential function of position, attractive delta
interactions in our problem. The time-independent Schr\"{o}dinger
equation on $M$ for the bound states of a particle under the
influence of $N$ attractive delta interactions reads
\be \label{scheq} \left[ \hbarm \triangle_g - \sum_{i=1}^N \; g_i
\; \delta^D(x-a_i) \right] \psi(x)= -\nu^2 \psi(x), \ee
where $g_{i} \in \mathbb{R}^+$ is the strength of the delta
interaction at $a_i$ and $-\nu^2$ is the bound state energy of the
system.
If we substitute (\ref{expansion1}) and (\ref{expansion2}) into
the Schr\"{o}dinger equation, it yields
\be \nonumber \sum_{l \geq 0} \left[ \hbarm \, \lambda_l \, C_l -
\sum_{i=1}^N A_i \, g_i \, \phi_l^*(a_i) + \nu^2 C_l \right]
\phi_l(x) = 0, \ee
where $A_i \equiv \psi(a_i)$ for simplicity of notation. The fact
that $\phi_l$'s form a complete orthonormal system allows us to
write $C_l$ in terms of them:
\be \label{cl} C_l = \frac{1}{\hbarm \, \lambda_l + \nu^2} \;
\sum_{i=1}^N A_i \, g_i \, \phi_l^*(a_i). \ee
Substituting (\ref{cl}) into the definition of $A_i$
\be \nonumber A_i = \sum_{j=1}^N A_j \, g_j \, \sum_{l \geq 0}
\frac{\phi_l(a_i) \phi_l^*(a_j)}{\hbarm \, \lambda_l + \nu^2} \, ,
\ee
and grouping the $A_i$ terms we find
\be \nonumber \label{phiA} \left[g_i^{-1} - \sum_{l \geq 0}
\frac{|\phi_l(a_i)|^2}{\hbarm \, \lambda_l + \nu^2} \right] A_i -
\sum_{\substack{j=1 \\
j \neq i}}^N \left[ \, \frac{g_j}{g_i} \, \sum_{l \geq 0}
\frac{\phi_l(a_i) \phi_l^*(a_j)}{\hbarm \, \lambda_l + \nu^2} \,
\right]A_j = 0 \;. \ee
The observation that the preceding equation is linear in $A_i$
permits us to write it naturally as a matrix equation
\be \label{PhiA=0}\Phi(-\nu^2)A = 0 \;,\ee
where $\Phi(-\nu^2)$ is called the characteristic matrix and
defined as
\be \label{phigeneral} \Phi_{ij}(-\nu^2) =
\begin{cases}
\begin{split}
g_i^{-1} - \sum_{l \geq 0} \frac{|\phi_l(a_i)|^2}{\hbarm \,
\lambda_l + \nu^2}
\end{split}
& \textrm{if $i = j$} \\ \\
\begin{split}
-\, \frac{g_j}{g_i} \, \sum_{l \geq 0} \frac{\phi_l(a_i)
\phi_l^*(a_j)}{\hbarm \, \lambda_l + \nu^2} \,
\end{split}
& \textrm{if $i \neq j$}\;.
\end{cases}
\ee
As we shall see below that the resolvent is intimately related to it and this allows us to state that the equation
$\det \Phi(-\nu^2)=0$ gives the bound state energies of our problem. In other words, this equation is considered to be
the determining equation of the ground state energy. Unfortunately, this nontrivial eigenvalue problem can not be
solved analytically, that is, we can not obtain an exact expression for the bound state energy for arbitrary $N$ since
the characteristic matrix depends \textit{nonlinearly} on the bound state energy. Indeed, the problem is even worse
than that, because we have not a finite expressions in the matrix elements of $\Phi_{ij}(-\nu^2)$. Fortunately, there
exist a way to redefine the problem so that the physical observables yield finite values with the help of
regularization and renormalization. Before introducing this procedure for our problem, it would be good to review first
the problem in flat spaces. The infinite sums in the characteristic matrix on $\mathbb{R}^2$ or $\mathbb{R}^3$ is then
replaced by integrals. The idea in that case is to take Fourier transform of the wave function
\be \nonumber \psi(x)= \int \tilde{\psi}(k)e^{i k.x} \frac{d^D
k}{(2\pi)^D}\;, \ee
and substitute into the Schr\"{o}dinger equation. Then we find
that the diagonal part of the characteristic matrix is
\be \nonumber \frac{1}{g_i}-\frac{1}{(2\pi)^D}\int \frac{d^D
k}{k^2 + \nu^2} \;, \ee
where $D=2,3$. This integral does not converge as it stands. The
well-known method to remove the divergence is to put a cut-off
$\Lambda$ to the integral's upper limit and consider the equation
as a determining equation of bound state energy for a given
coupling constant $g$. If this regularization is performed, we
realize that as the cut-off goes to infinity, ground state energy
becomes divergent. In order to get a physically acceptable result,
one assumes that the coupling constant depends on this cut-off and
performs the limit $\Lambda\rightarrow \infty$ in such a way that
bound state energy remains finite. These infinities should be
removed properly since all the physical observables are measured
experimentally as finite quantities. The cut-off dependence of the
coupling constant is chosen as
\be \frac{1}{g_{i}(\Lambda)}=\frac{1}{(2\pi)^D}\int_{|k|<\Lambda}
\frac{d^D k}{k^2 + \mu_i^2} \;.\ee
The determination of this coupling constant is called renormalization. Now, we follow the same idea to remove the
divergence from our problem. By using Weyl's asymptotic formula \cite{Weyl}, one expects that the diagonal term
$\sum_{l \geq 0} \frac{|\phi_l(a_i)|^2}{\hbarm \, \lambda_l + \nu^2}$ in the above matrix does not converge and this
will be explicitly seen for a particular manifold $\mathbb{S}^2$. For a general compact manifold, we introduce cut-off
to the upper bound of the infinite sum and choose the coupling constant as
\be g_{i}^{-1}(\Lambda)= \sum_{l = 0}^{\Lambda}
\frac{|\phi_l(a_i)|^2}{\hbarm \, \lambda_l + \mu_{i}^2} \;, \ee
where $-\mu_{i}^{2}$ is the measured binding energy to a single
delta interaction. Then, we take the limit $\Lambda \rightarrow
\infty$
\be \label{comprenorm} \lim_{\Lambda\rightarrow\infty}
\left[\sum_{l = 0}^{\Lambda} \frac{|\phi_l(a_i)|^2}{\hbarm \,
\lambda_l + \mu_{i}^2} - \sum_{l = 0}^{\Lambda}
\frac{|\phi_l(a_i)|^2}{\hbarm \, \lambda_l + \nu^2}\right]\;,\ee
and this should give us a finite result in two and three dimensions. Hence, the divergence has been removed and bound
state energy becomes finite. A rigorous proof of this is not trivial, so we will stay at a heuristic level and study
special cases only.

As we will show in the next subsection, the heat kernel is intimately related to the characteristic matrix $\Phi$ and
this relation helps us to see easily which part of the matrix is divergent or convergent and then how to renormalize
the problem non-perturbatively. Furthermore, heat kernel is especially very helpful to remove the divergences for our
problem on non-compact manifolds, as we shall discuss in section \ref{delta on hyperbolic}. We will see that the above
method can easily be extended to find the renormalized resolvent of the singular Hamiltonian.

\subsection{\label{phi and heat kernel} The Relation of Matrix $\Phi$ with Heat Kernel and Resolvent}

The resolvent (or Green's function) and heat kernel play very
essential role in establishing the connection between spectral
properties of the operator and corresponding geometrical notions.
Up to now, we have been dealing with a matrix $\Phi$, and do not
refer to resolvent and heat kernel. In order to see the relation
between the matrix $\Phi$ and heat kernel we consider the
separable Hamiltonians $H = H_0 - \sum_{i=1}^N g_i | f_i \rangle
\langle f_i |$, where $|f_i\rangle$ is a particular Dirac ket. We
work out the resolvent formula of $H$ in terms of $H_0$ and assume
that the two Dirac kets $| \psi \rangle$ and $| \chi \rangle$ are
related in such a way that the equality $(H-z) | \psi \rangle = |
\chi \rangle$ is satisfied. Then, we have
\be \label{reseq} \left[ H_0 - z - \sum_{j=1}^N g_j | f_j \rangle
\langle f_j | \right] | \psi \rangle = | \chi \rangle \;, \ee
assuming complex number $z \not \in \mathrm{Spec}(H_{0})$. Acting
the operator $(H_0 - z)^{-1}$ on both sides and projecting it onto
$\langle f_i |$, we obtain
\be \nonumber \label{phifi} \sum_{j = 1}^N \Phi_{ij}(z) \langle
f_j | \psi \rangle = g_i^{-1} \, \langle f_i |\left( H_0 - z
\right)^{-1} | \chi \rangle \;, \ee
where we define a matrix $\Phi_{ij}(z)$ as \footnote{There is no
confusion in notation because we will see that this matrix $\Phi$
is exactly the same matrix considered in the previous sections.}
\be \label{phiheat} \Phi_{ij} (z) =
\begin{cases}
\begin{split}
g_i^{-1} - \langle f_i | \left( H_0 - z \right)^{-1}|f_i \rangle
\end{split}
& \textrm{if $i = j$} \\
\begin{split}
- \frac{g_j}{g_i} \; \langle f_i |\left( H_0 - z \right)^{-1} |
f_j \rangle
\end{split}
& \textrm{if $i \neq j$}.
\end{cases}
\ee
After a little algebra, it is evident that
\be \left( H - z \right)^{-1} = \left( H_0 - z \right)^{-1} +
\left( H_0 - z \right)^{-1} \left[ \sum_{i,j=1}^N | f_i \rangle
\Phi_{ij}(z)^{-1} \langle f_j | \right] \left( H_0 - z
\right)^{-1}\;, \label{aij} \ee
as long as  $\Phi_{ij}(z)^{-1}$ exists. Such formulae were extensively discussed in problems associated with
self-adjoint extensions of operators, notably by Krein and his school, and also for such singular interactions in flat
spaces \cite{Albeverio 2004, Albeverio 2000}. Therefore, our problem can also be considered  as a kind of self-adjoint
extension of the free Hamiltonian. It is defined through regulating (or controlling) the behavior of the wave function
in the vicinity of these interaction points.

If we take the matrix element of (\ref{aij}) by projecting on to
the Dirac kets $\langle x |$ and $|y \rangle$, we have found the
resolvent kernel $R(x,y|z) \equiv \langle x | (H-z)^{-1}|y \rangle
$ corresponding to (\ref{reseq})
\be \nonumber
\begin{split}
& R(x,y|z) = R_0(x,y|z) + \int dx' dy' R_0(x,x'|z) \left[
\sum_{i,j=1}^N f_i(x') \Phi_{ij}(z)^{-1} f_j(y') \right] R_0(y',y|z)  \\
&= R_0(x,y|z) + \sum_{i,j=1}^N   \left[ \int dx' R_0(x,x'|z)
f_i(x') \right] \Phi_{ij}(z)^{-1} \left[ \int dy' R_0(y',y|z)
f_j(y') \right].
\end{split}
\ee
By choosing the functions $f_i(x)$'s as bump functions centered at
$x=a_i$ such that the sequences of the functions admit the limit
$f_i(x) \rightarrow \delta^D(x-a_i)$ (in the appropriate
topology), it turns out that
\be \label{resolvent} R(x,y|z) = R_0(x,y|z) + \sum_{i,j=1}^N
R_0(x,a_i|z)\, \Phi_{ij}(z)^{-1} R_0(a_j,y|z)\;. \ee
The important point to note here is the relation between the
resolvent operator, defined on an infinite dimensional space and
the characteristic matrix, defined on a finite dimensional space.
This allows us to find the bound state spectrum of the separable
Hamiltonian operator $H$ with the help of a finite dimensional
matrix $\Phi(z)^{-1}$. Since discrete spectrum is the set of
complex numbers such that the resolvent does not exist, this
proves that the equation $\det \Phi =0$ gives the bound state
spectrum of our system. The fact that the free Hamiltonian is
bounded from below allows us to write the free resolvent operator
as an integral for $\Re(z) < 0$
\be \left( H_0 - z \right)^{-1} = \frac{1}{\hbar} \int_0^{\infty}
e^{-\frac{t}{\hbar} \left(\hbarm \, \triangle_g - z \right)} dt,
\ee
the result of which should be continued analytically to its
largest set in the entire complex plane. As a consequence of this,
the free resolvent kernel is
\be \nonumber \langle f_i | \left( H_0 - z \right)^{-1}|f_j
\rangle =\frac{1}{\hbar} \int_0^{\infty} e^{\frac{z t}{\hbar}}
\langle f_i|\, e^{-\left[\frac{t}{\hbar}\right] \hbarm \,
\triangle_g}|f_j \rangle \, dt. \ee
Taking the limit $f_i(x) \rightarrow \delta^D(x-a_i)$, it results
in
\be R_{0}(a_i, a_j |z) = \langle a_i | \left( H_0 - z
\right)^{-1}|a_j \rangle = \frac{1}{\hbar}\int_0^{\infty}
e^{\frac{z t}{\hbar}} K_t(a_i,a_j) \, dt\;, \label{heatkernel and
A} \ee
where $K_t(a_i, a_j)$ is the so-called heat kernel, and the
operator $e^{-\left[\frac{t}{\hbar}\right] \left(\hbarm \,
\triangle_g\right)}$ is the formal solution to the heat equation
\cite{Rosenberg,Davies}. Hence, the matrix $\Phi$ is written in
terms of the heat kernel in the following way,
\be \label{heatkernel,phi} \Phi_{ij}(z) =
\begin{cases}
\begin{split}
g_i^{-1} - \frac{1}{\hbar}\int_0^{\infty} e^{\frac{z t}{\hbar}}
K_t(a_i,a_i) \, dt
\end{split}
& \textrm{if $i = j$} \\
\begin{split}
- \frac{g_j}{g_i} \; \frac{1}{\hbar}\int_0^{\infty} e^{\frac{z
t}{\hbar}} K_t(a_i,a_j) \, dt
\end{split}
& \textrm{if $i \neq j$}.
\end{cases}
\ee
The matrix $\Phi_{ij}$ is exactly the same matrix mentioned in the
previous sections. This can be shown easily from the spectral
theorem \cite{Rosenberg} for compact manifolds
\be K_t(a_i,a_j) = \sum_{l \geq 0} e^{- \hbarm \lambda_l
\left[\frac{t}{\hbar}\right] } \phi_l(a_i) \phi_l^*(a_j), \ee
which converges uniformly on $M \times M$ for each $t > 0$:
\be \nonumber \label{esums}
\begin{split}
\langle f_i | \left( H_0 - z \right)^{-1}|f_j \rangle &\rightarrow
\int_0^{\infty} e^{\frac{z t}{\hbar}} K_t(a_i,a_j) \,
 \frac{dt}{\hbar}
= \sum_{l \geq 0} \phi_l(a_i) \phi_l^*(a_j) \int_0^{\infty}
e^{-\left(\hbarm \lambda_l - z\right)\left[\frac{t}{\hbar}\right]}
\, \frac{dt}{\hbar} \\
&= \sum_{l \geq 0} \frac{\phi_l(a_i) \phi_l^*(a_j)}{\hbarm
\lambda_l - z} \;,
\end{split}
\ee
where summation and integral are interchanged since summation
converges uniformly. This is the same result for $z=-\nu^2$ that
we already obtained for non-diagonal part of the characteristic
matrix in the section \ref{rencompact}. One can understand how the
non-diagonal part of it in (\ref{phigeneral}) is convergent by
using the smooth behaviour of the heat kernel and the integral
$\int_0^{\infty} e^{\frac{z t}{\hbar}} K_t(a_i,a_j) $ is
convergent for $a_i \neq a_j$. However, the asymptotic behaviour
of the heat kernel as $t\rightarrow 0^+$ for every point $x$ on a
compact manifold  $M$ \cite{Rosenberg} is given by
\be K_t (x,x) \sim  \left(4\pi \frac{\hbar t}{2m}\right)^{-D/2}
\sum_{k=0}^{\infty} u_{k}(x,x) \left(\frac{\hbar t}{2m}\right)^k
\;, \ee
where $D$ is the dimension of the manifold and the $u_k (x,x)$ are
functions given in terms of the curvature tensor of $M$ and its
covariant derivatives at the point $x$. This result shows that
diagonal part of the heat kernel as $t\rightarrow 0^+$ for $D=2,3$
leads to a divergence since $u_0 (x,x)=1$ (there is no infinities
for $D=1$ as it can be easily realized). In other words, the sum
in the diagonal term in $\Phi$ is divergent while the sum in the
non-diagonal term is convergent. However, we have already shown
that bound state energies are related to the characteristic
matrix, i.e., $\det \Phi(z) =0$ contains information about bound
states. If some of the elements of the characteristic matrix have
infinities, it is impossible to get sensible bound state energies
for our problem. Before establishing the renormalization of our
problem with the help of heat kernel, we must indicate why this
problem occurs. Although the delta interactions may approximately
describe a system in which a particle interacting with a
point-like centers when its de Broglie wavelength is large
compared to the typical range of a potential, we have not
encountered in nature this type of contact interaction. This means
that the substituting the Dirac-Delta interactions into the
Hamiltonian for $D=2,3$ directly is not a proper way. Therefore,
we must modify our problem such that it has a finite range and
then consider the zero range limit. In our renormalization method
with heat kernel, short range is replaced with the short time as
we will see.

We introduce a small constant $\epsilon$, in the lower limit of
the integral. We then take the limit as the cut-off $\epsilon$
goes to zero in such a way that the experimentally measured ground
state energy remains finite. This requires that some quantities in
the problem, e.g. coupling constant, should have a cut-off
dependence in a definite way. For our problem, we naturally choose
\be g_i^{-1}(\epsilon,\mu_i) =
\frac{1}{\hbar}\int_\epsilon^{\infty} e^{\frac{-\mu_i^2 t}{\hbar}}
K_t(a_i,a_i) \, dt \;. \ee
After performing the limit $\epsilon \rightarrow 0$, we have the
renormalized characteristic matrix
\be \label{heatkernel,phi} \Phi_{ij}(z) =
\begin{cases}
\begin{split}
\frac{1}{\hbar} \int_0^{\infty} K_t(a_i,a_i)
\left[e^{\frac{-\mu_{i}^{2} t}{\hbar}} - e^{\frac{z t}{\hbar}}
\right]\, dt
\end{split}
& \textrm{if $i = j$} \\
\begin{split}
- \frac{1}{\hbar}\int_0^{\infty} e^{\frac{z t}{\hbar}}
K_t(a_i,a_j) \, dt
\end{split}
& \textrm{if $i \neq j$}\;,
\end{cases}
\ee
where $\Re(z)<0$ and $ \Phi_{ij}(z)$ can be analytically continued
to its largest set in the entire complex plane. One can naturally
ask whether the renormalization performed with heat kernel is
compatible with the one introduced in section \ref{rencompact}.
The answer is affirmative and one can easily show that the cut-off
$\Lambda$ for the infinite sum introduced in section
\ref{rencompact} corresponds to the cut-off $\epsilon$ for the
lower bound of integral in the heat kernel method. This can be
realized easily by using the spectral theorem in the diagonal part
of equation (\ref{heatkernel,phi}) and taking $z=-\nu^2$:
\be
\begin{split}
& g_i^{-1}(\epsilon,\mu_i) - \frac{1}{\hbar}\int_\epsilon^{\infty}
e^{\frac{-\nu^2 t}{\hbar}} \sum_{l \geq 0} e^{- \hbarm \lambda_l
\left[\frac{t}{\hbar}\right] } \phi_l(a_i) \phi_l^*(a_i)\, dt \\ =
& g_i^{-1}(\epsilon,\mu_i) - \frac{1}{\hbar}\sum_{l \geq
0}\phi_l(a_i) \phi_l^*(a_i)\int_\epsilon^{\infty} e^{\frac{-\nu^2
t}{\hbar}}  e^{- \hbarm \lambda_l \left[\frac{t}{\hbar}\right] }
\, dt \;,
\end{split}
\ee
where we have used the uniform convergence of the sum. Now, in
order to remove the divergence, we can naturally choose the
coupling constant as
\be g_i^{-1}(\epsilon,\mu_i) = \frac{1}{\hbar}\sum_{l \geq
0}\phi_l(a_i) \phi_l^*(a_i)\int_\epsilon^{\infty}
e^{\frac{-\mu_{i}^{2} t}{\hbar}}  e^{- \hbarm \lambda_l
\left[\frac{t}{\hbar}\right] } \, dt \;. \ee
Then, we have
\be
\begin{split} &
\lim_{\epsilon\rightarrow 0} \left\{\frac{1}{\hbar}\sum_{l \geq
0}\phi_l(a_i) \phi_l^*(a_i) \left[\int_\epsilon^{\infty}
e^{\frac{-\mu_{i}^{2} t}{\hbar}}  e^{- \hbarm \lambda_l
\left[\frac{t}{\hbar}\right] } \, dt - \int_\epsilon^{\infty}
e^{\frac{-\nu^2 t}{\hbar}}  e^{- \hbarm \lambda_l
\left[\frac{t}{\hbar}\right]} \, dt \right] \right\}
\\ = & \frac{1}{\hbar}\sum_{l \geq 0}
\frac{\left[\nu^2 - \mu_{i}^2\right] \phi_l(a_i)
\phi_l^*(a_i)}{\left[\hbarm \lambda_l +
\mu_{i}^2\right]\left[\hbarm \lambda_l + \nu^2\right]} \;,
\end{split}
\ee
which is the same result we would have obtained by the
eigenfunction expansion by introducing a cut-off $\Lambda$
(equation (\ref{comprenorm})). After finding the renormalized
characteristic matrix, the resolvent can be written explicitly
\be R(x,y|z) = R_0(x,y|z) + \sum_{i,j=1}^N R_0(x,a_i|z)\,
\Phi_{ij}(z)^{-1} R_0(a_j,y|z)\;, \ee
where
\be R_{0}(x,y |z) = \frac{1}{\hbar}\int_0^{\infty} e^{\frac{z
t}{\hbar}} K_t(x,y) \, dt\;. \ee
Once we have given the resolvent of an operator, all the information about the operator is contained in it.
Nevertheless, it is instructive to check that the wave functions can also be obtained and they are normalizable. We
write the normalized wave function with a cut-off $\Lambda$ and then take the limit $\Lambda \to \infty$, this way we
will not get a vanishing wave function. So the normalization constant can be found easily
\begin{eqnarray}
\begin{split}
|C(\Lambda)|^{-2} &= \sum_{i,j=1}^N g_i(\Lambda) \, g_j(\Lambda)
\, A_i^{*}(\Lambda) \, A_j(\Lambda) \int d^D x \sqrt{g}
\sum_{l,l'=0}^{\Lambda} \frac{\phi_l(a_i)
\phi_l^*(x)}{\left(\hbarm \, \lambda_l + \nu^2 \right)}
\frac{\phi_{l'}^{*}(a_j) \phi_{l'}(x)}{\left(\hbarm \,
\lambda_{l'} +
\nu^2 \right)}\\
&=\sum_{i,j=1}^N g_i(\Lambda) \, g_j(\Lambda) \, A_i^{*}(\Lambda)
\, A_j(\Lambda) \sum_{l=0}^{\Lambda} \frac{\phi_l(a_i)
\phi_l^*(a_j)}{\left(\hbarm \, \lambda_l + \nu^2 \right)^2} \;.
\end{split}
\end{eqnarray}
One expects from the Weyl asymptotic formula that the wave
function is not normalizable if we are on a space of dimension
bigger than three. Moreover, we can see that the summation over
the eigenmodes is exactly the derivative of $\Phi(-\nu^2)$ with
respect to $\nu$, hence we get:
\be |C(\Lambda)|^{-2}= \frac{1}{2\nu} \sum_{i,j=1}^N
g_j(\Lambda)^{2} A_i^{*}(\Lambda) \; \frac{\partial
\Phi_{ij}(\Lambda, -\nu^2)}{\partial \nu} \, A_j(\Lambda). \ee
Performing the limit $\Lambda \rightarrow \infty$, the properly
normalized wave function of n$^{th}$ state becomes \be \nonumber
\begin{split}
\psi_n(x) = \sqrt{2 \nu_n} & \left[ \sum_{r,s=1}^N A_r^{*}(\nu_n)
\, \left.\frac{\partial \Phi_{r s}(-\nu^2)}{\partial
\nu}\right|_{\nu = \nu_n} A_s(\nu_n) \right]^{-\frac{1}{2}} \;
\sum_{l \geq 0} \sum_{i=1}^N A_i(\nu_n) \; \frac{\phi_l^{*}(a_i)
\phi_l(x)}{\left(\hbarm \, \lambda_l + \nu_{n}^2 \right)} \;,
\end{split}
\ee
where $\nu_n$ is the n$^{th}$ root of the energy equation $\det
\Phi(-\nu^2) = 0$. This can further be simplified to an expression
in terms of the heat kernel
\be
\begin{split}
\psi_n(x) = \sqrt{2 \nu_n} & \left[ \sum_{r,s=1}^N A_r^{*}(\nu_n)
\, \left.\frac{\partial \Phi_{r s}(-\nu^2)}{\partial
\nu}\right|_{\nu = \nu_n} A_s(\nu_n)
\right]^{-\frac{1}{2}} \; \\
& \qquad \qquad \qquad \qquad \qquad \times \sum_{l \geq
0}\sum_{i=1}^N A_i(\nu_n) \; \phi_l^{*}(a_i)
\phi_l(x)\int_{0}^{\infty} e^{-\frac{t}{\hbar} \left(\hbarm \, \lambda_l + \nu_{n}^2 \right)} \frac{dt}{\hbar} \\
= \sqrt{2 \nu_n} & \left[ \sum_{r,s=1}^N A_r^{*}(\nu_n) \, \left.\frac{\partial \Phi_{r s}(-\nu^2)}{\partial
\nu}\right|_{\nu = \nu_n} A_s(\nu_n) \right]^{-\frac{1}{2}} \; \int_{0}^{\infty} e^{-\frac{t \nu_{n}^2}{\hbar}}
\sum_{i=1}^N A_i(\nu_n) K_{t}(a_i, x) \frac{dt}{\hbar} \;,
\end{split}
\ee
in which one can easily see that $\psi_n(x)$ is finite.

\section{\label{Delta Interactions on Sphere $S^2$} Finitely Many Dirac-Delta Interactions on $\mathbb{S}^2$}

Since the simplest and one of the most familiar compact manifolds
is the sphere $\mathbb{S}^2$, we shall work out the problem of
point interactions on a sphere as a concrete example. Suppose that
point interactions are located at the points given by the local
coordinates $(\theta_i, \phi_i)_{i=1}^N$ on a sphere of radius
$R$. Then, the Schr\"{o}dinger equation for the bound states of a
particle living on the sphere under the influence of $N$
attractive delta interactions becomes
\be \left[\hbarm \triangle_{\mathbb{S}^2} - \sum_{i=1}^N \; g_i \;
\delta^2(\theta-\theta_i, \phi-\phi_i)  \right] \psi = -\nu^2
\psi, \ee
where $\triangle_{\mathbb{S}^2}$ is Laplacian on the sphere in
spherical coordinates
\be \triangle_{\mathbb{S}^2} = - \frac{1}{R^2 \sin
\theta}\frac{\partial}{\partial \theta}\left(\sin \theta
\frac{\partial} {\partial \theta}\right) - \frac{1}{R^2 \sin^2
\theta} \frac{\partial^2}{\partial\phi^2}\;, \ee
and $\delta^2(\theta-\theta_i,
\phi-\phi_i)=\frac{\delta(\theta-\theta_i)\,\delta(
\phi-\phi_i)}{R^2 \sin^2 \theta}$ is the two dimensional delta
function on the sphere centered at $(\theta_i, \phi_i)$. It is
well known that spherical harmonics $Y_{l}^{m}$ are eigenfunctions
of the Laplacian $\triangle_{\mathbb{S}^2}$ with the eigenvalues
$l(l+1)/R^2$ and form a complete orthonormal basis on
$\mathbb{S}^2$. In order to be consistent with the standard
normalization of spherical harmonics, we choose
$\phi_{lm}=\frac{Y_{l}^{m}}{R}$. From the following identity
\bea \sum_{m = -l}^l Y_l^m(\theta_i, \phi_i) {Y_l^m}^*(\theta_j,
\phi_j) & = & \frac{2l+1}{4 \pi} \; P_l \left(\cos \theta_i \cos
\theta_j + \cos(\phi_i - \phi_j) \sin \theta_i \sin \theta_j
\right) \nonumber \\\nonumber & = & \frac{2l+1}{4 \pi} \; P_l
\left( 1 - \frac{\mathrm{d}_{ij}^{2}}{2} \right)\;, \eea
where $\mathrm{d}_{ij} = \frac{d_{ij}}{R} = |\hat{r}_i -
\hat{r}_j| \in [0,2]$ being rescaled distance between point
centers with radius of the sphere $R$, the matrix
$\Phi_{ij}(-\nu^2)$ in (\ref{phigeneral}) becomes
\be \label{phi1} \Phi_{ij}(-\nu^2) =
\begin{cases}
\begin{split}
g_i^{-1} - \frac{1}{4 \pi R^2}\sum_{l \geq 0} \frac{2l+1}{\hbarmr
l(l+1) + \nu^2}
\end{split}
& i = j \\ \\
\begin{split}
-\frac{g_j}{g_i}\frac{1}{4 \pi R^2}\sum_{l \geq 0}
\frac{2l+1}{\hbarmr l(l+1) + \nu^2} \; P_l\left( 1 -
\frac{\mathrm{d}_{ij}^2}{2}\right)
\end{split}
& i \neq j.
\end{cases}
\ee
It follows easily from the Cauchy-MacLaurin integral test that the
infinite sum
\be \nonumber \frac{1}{4\pi R^2}\sum_{l \geq 0}\frac{2l+1}{\hbarmr
\; l(l+1) + \nu^2} \ee
is divergent. To get a sensible results for our problem, we must
modify our original problem as outlined in section
\ref{rencompact}. Therefore, considering our problem in the light
of this method, we first define the coupling constant $g_i$ as a
function of the parameter $\Lambda$ (cut-off). Then, by choosing
$g_i^{-1}(\Lambda)$'s naturally
\be \nonumber g_i^{-1}(\Lambda) = \frac{1}{4 \pi R^2} \sum_{l =
0}^\Lambda \frac{2l+1}{\hbarmr \; l(l+1) + \mu_i^2} \;, \ee
where $\mu_i$ is experimentally measured value of bound state
energy for the single delta interaction and taking the limit
$\Lambda \rightarrow \infty$ of the difference, we have obtained
\bea \nonumber && \lim_{\Lambda \rightarrow \infty}\left[
\frac{1}{4 \pi R^2} \sum_{l = 0}^\Lambda \frac{2l+1}{\hbarmr \;
l(l+1) + \mu_i^2}-\frac{1}{4\pi
R^2}\sum_{l=0}^{\Lambda}\frac{2l+1}{\hbarmr \; l(l+1) +
\nu^2}\right]\\\nonumber&& \qquad  \qquad \qquad \longrightarrow
\frac{1}{4 \pi R^2 \mu_R^2} \left[\phi \left( \frac{\mu_i}{\mu_R}
\right)-\phi \left( \frac{\nu}{\mu_R} \right)\right]\;, \eea
where $\mu_{R}^{2}\equiv \hbarmr$. The function $\phi$ here is
defined as
\be \nonumber \phi(x) \equiv
\frac{1}{x^2}-H_{\frac{1}{2}-\sqrt{\frac{1}{4}-x^2}}-H_{\frac{1}{2}+\sqrt{\frac{1}{4}-x^2}}
\;, \hspace{2cm}x \in \mathbb{R}^{+}\;, \ee
where $H$'s are the harmonic numbers, commonly defined on integers
as $H_n =\sum_{k=1}^{n}\frac{1}{k}$ and can be extended by
analytical continuation to its largest domain in the entire
complex plane as $H_z =\psi(z+1)+\gamma$, where
$\psi(z)=\frac{\Gamma'(z)}{\Gamma (z)}$ being the digamma function
and $\gamma$ being the Euler-Mascheroni constant. The digamma
function has several useful integral representations
\cite{specialfunc}, some of which are
\bea \label{digammarep1}
\psi(z) &=& \int_{0}^{\infty} \left(\frac{e^{-t}}{t}-\frac{e^{-z t}}{1-e^{-t}}\right)dt \;,\\
\psi(z) &=& \log z + \int_{0}^{\infty}
\left(\frac{1}{1-e^{-t}}+\frac{1}{t}-1 \right) e^{-z t}dt \;,
\label{digammarep2} \eea
where $\Re(z)>0$ and these can be useful for the estimates of its
upper and lower bounds. Due to the Schwarz reflection principle of
harmonic numbers ($\bar{H_z}=H_{\bar{z}}$), the function $\phi(x)$
is real valued ($\phi \in \mathbb{R}$) for all $x\in
\mathbb{R}^{+}$. It is also easy to check $\lim_{\Lambda
\rightarrow \infty}
\frac{g_{j}(\Lambda)}{g_{j}(\Lambda)}\rightarrow 1$ in the
non-diagonal part of (\ref{phi1}) , simply because of their same
form of the divergence. Then, the renormalized matrix
$\Phi(-\nu^2)$ for bound states can be eventually written as
\be \label{phi} \Phi_{ij}(-\nu^2) = \frac{1}{4 \pi R^2 \mu_R^2}
\begin{cases}
\begin{split}
 \phi \left(
\frac{\mu_i}{\mu_R} \right)-\phi \left( \frac{\nu}{\mu_R} \right)
\end{split}
& i = j \\ \\
\begin{split}
-\sum_{l \geq 0} \frac{2l+1}{l(l+1) + \frac{\nu^2}{\mu_R^2}} \;
P_l\left( 1 - \frac{\mathrm{d}_{ij}^2}{2}\right)
\end{split}
& i \neq j \;.
\end{cases}
\ee
By the analytical continuation of the characteristic matrix to its
largest domain in the entire complex plane, we have
\be \label{phi} \Phi_{ij}(z) = \frac{1}{4 \pi R^2 \mu_R^2}
\begin{cases}
\begin{split}
 \phi \left(
\frac{\mu_i}{\mu_R} \right)-\phi \left( \frac{\sqrt{-z}}{\mu_R}
\right)
\end{split}
& i = j \\ \\
\begin{split}
-\sum_{l \geq 0} \frac{2l+1}{l(l+1) - \frac{z}{\mu_R^2}} \;
P_l\left( 1 - \frac{\mathrm{d}_{ij}^2}{2}\right)
\end{split}
& i \neq j \;,
\end{cases}
\ee
from which we can write the resolvent equation (\ref{resolvent}).
Hence, we have obtained a well-defined formulation of our problem,
that is, the infinities have been removed. Moreover, we  see that
the problem realizes a generalized dimensional transmutation. In
this case, the coupling constants $g_i$ have the same dimension as
$\hbarm$ by dimensional analysis. In contrast to the flat case, we
have one more parameter $R$ coming from the geometry of the space.
Thus, we expect that the system must have an  intrinsic energy
scale $\hbarmr$ as well as $\frac{\hbar^2}{m d_{ij}^2}$ terms.
However, after the renormalization, we obtain a set of new
dimensional parameters $\mu_i^2$. Hence, the first set of scales
we expect by naive dimensional analysis at the beginning is not
sufficient. Instead, a specific combination of all these
parameters together determine the scale of our problem. This means
that delta potentials on a sphere is an example of a kind of
dimensional transmutation. However, there is a slight difference,
especially in the case of single delta attractor: in the flat case
there is no combination of dimensional parameters to come up with
an energy scale, whereas in the case of a sphere we have a
geometric length scale  $R$ which already defines an energy scale
$\hbarmr$. The dimensional transmutation is most striking in such
cases where there is no intrinsic energy scale.

In order to estimate the non-diagonal part of the matrix $\Phi$
for sphere $\mathbb{S}^2$, we follow a different strategy, using
the heat kernel.

\subsection{\label{heatkernel for S^2} Lower Bound of $ E_{gr}$ by Heat Kernel Method for $\mathbb{S}^2$}

Heat kernel $K_t(x,y)$ is the unique fundamental solution to the
heat equation $\frac{\hbar^2}{2m}\triangle_{g} \phi = - \hbar
\phi_t$. It has the symmetry ($K_t(x,y) = K_t(y,x)$) and
semi-group property \cite{Rosenberg,Davies}. As well as being a
useful computational tool in establishing the existence and some
of the properties of the spectrum of the Laplacian of the
eigenfunctions on a Riemannian manifolds, it is very helpful to
understand the nature of the divergences for our purposes, as we
have shown in the previous section.

By means of the relation (\ref{heatkernel,phi}) and explicit form
of the heat kernel, one can calculate the matrix $\Phi_{ij}$.
However, there are some situations in which one can not calculate
the heat kernel explicitly, e.g. we do not have  an explicit
expression of the heat kernel for two dimensional sphere. In this
case, one can still find some bound estimates on matrix
$\Phi_{ij}$ without having explicit form of the heat kernel,
instead some properties of it. In order to analyze this for
$\mathbb{S}^2$, we will use some estimates on the heat kernel,
based on a work by Li and Yau \cite{liyau par}. Let us recall the
corollary of the theorem (3.1) in \cite{liyau par}:

Let $M$ be a complete manifold without boundary. If the Ricci
curvature of $M$ is bounded from below by $-K$, for some constant
$K\geq 0$, then for $1< \alpha < 2$ and $0< \varepsilon < 1$, the
heat kernel satisfies
\be \nonumber K_t (x,y) \leq
\frac{C(\varepsilon)^{\alpha}}{\sqrt{V(x,\sqrt{t})V(y,\sqrt{t})}}
\;e^{C_{7}\, \varepsilon (\alpha -1)^{-1}K t -
\frac{d(x,y)^{2}}{(4+\varepsilon) t}}\;, \ee
where $V(x,r)= \mu (B(x,r))$, $B(x,r)$ is the geodesic ball of
radius $r$ centered at $x \in M$ and $d(x,y)$ is the geodesic
distance between two points $x$ and $y$ on the manifold. The
constant $C_{7}$ depends only on the dimension of the manifold
$D$, while $C(\varepsilon)$ depends on $\varepsilon$ with
$C(\varepsilon) \rightarrow \infty$ as $\varepsilon \rightarrow
0$. When $K=0$, the above estimate, after letting $\alpha
\rightarrow 1$, can be written as
\be  K_t (x,y) \leq
\frac{C(\varepsilon)}{\sqrt{V(x,\sqrt{t})V(y,\sqrt{t})}} \;e^{-
\frac{d(x,y)^{2}}{(4+\varepsilon) t}}\;. \ee
Since $\mathbb{S}^2$ satisfies the conditions above as a
particular case, this corollary can be applied to it as well. On
the other hand, we have a different purpose from the original
corollary of the theorem for the estimates on the upper bound of
the heat kernel, in which the sharp estimate for the heat kernel
is found. Instead, we are trying to find a best lower bound of the
ground state energy of the system. Therefore, we shall modify the
original corollary in \cite{liyau par}. Using this theorem with
relaxed condition $0 < \varepsilon < 1$, we have found the upper
bound estimate for heat kernel of sphere $\mathbb{S}^2$ in our
problem:
\be K_t (a_i, a_j)\leq \frac{C'(\delta)}{\sqrt{V
\left(x,\sqrt{\frac{\hbar t}{2 m }}\right)V
\left(y,\sqrt{\frac{\hbar t}{2 m }}\right)}} \;e^{-\frac{2 m
d_{ij}^{2}}{D(\delta) \hbar t}}\;, \ee
where
\be C'(\delta)\equiv(1+\delta)^2 \exp \left[\frac{1}{4 \delta(1 +
\delta)(1 + 2\delta)}+\frac{1}{2\delta (2+\delta)}+
\frac{1}{4\delta}\right] \;, \ee
and \be D(\delta)\equiv 4(1+2\delta)(1+\delta)^2 \;, \ee
$\delta$ is merely required to be positive. When we want to find a lower bound for the energy, the numerical values of
the coefficients $C'(\delta)$ and $D(\delta)$ will be determined explicitly.
It is easy to see that $V\left(x,\sqrt{\frac{\hbar t}{2 m
}}\right)= V \left(y,\sqrt{\frac{\hbar t}{2 m }}\right)= 2 \pi R^2
\left(1-\cos \sqrt{\frac{\hbar t}{2 m R^2}}\right)$ as long as $0
\leq t \leq \frac{2 m \pi^2 R^2}{\hbar}$. For $t \geq \frac{2 m
\pi^2 R^2}{\hbar}$, we have $V\left(x,\sqrt{\frac{\hbar t}{2 m
}}\right)=V \left(y,\sqrt{\frac{\hbar t}{2 m }}\right)= 4 \pi
R^2$. According to our corollary and positive definiteness of heat
kernel, the following integral has an upper bound:
\be
\begin{split}
\nonumber \label{int bound kernel} &
\frac{1}{\hbar}\int_0^{\infty} e^{-\frac{\nu^2 t}{\hbar}}
K_t(a_i,a_j) \, dt \\ & \qquad \qquad  \qquad  \leq
\frac{C'(\delta)}{\hbar} \int_{0}^{\frac{2 m \pi^2 R^2}{\hbar}}
\frac{e^{-\frac{2 m d_{ij}^{2}}{D(\delta) \hbar t}-\frac{\nu^2
t}{\hbar }}}{2 \pi R^2 \left(1- \cos \sqrt{\frac{\hbar t}{2 m
R^2}}\right)} \;dt + \frac{C'(\delta)}{4\pi R^2 \hbar}
\int_{\frac{2 m \pi^2 R^2}{\hbar}}^{\infty}\; e^{-\frac{2 m
d_{ij}^{2}}{D(\delta) \hbar t}-\frac{\nu^2 t}{\hbar }}\;\;dt \;,
\end{split}
\ee
where we have taken $z$ as $-\nu^2$. With the help of the identity
$1-\cos \sqrt{\frac{\hbar t}{2 m R^2}}= 2 \sin^2 \sqrt{\frac{\hbar
t}{8 m R^2}}$ and the inequality $\frac{1}{\sin \theta} \leq
\frac{\pi}{2 \theta}$ for $0 \leq \theta \leq \pi/2$, we obtain
\bea \nonumber \frac{1}{\hbar} \int_0^{\infty} e^{-\frac{\nu^2
t}{\hbar}} K_t(a_i,a_j) \, dt & \leq &  \frac{m \pi C'(\delta)}{2
\hbar^2} \int_{0}^{\infty} \frac{e^{-\frac{2 m d_{ij}^{2}}
{D(\delta) \hbar t}-\frac{\nu^2 t}{\hbar }}}{t}\, dt +
\frac{C'(\delta)}{4\pi R^2 \hbar} \int_{0}^{\infty}\; e^{-\frac{2
m d_{ij}^{2}}{D(\delta) \hbar t}- \frac{\nu^2 t}{\hbar }}\, dt \;.
\eea
Evaluating these integrals, we find
\be
\begin{split}
\nonumber \label{phi ij} |-\Phi_{i\neq j}(-\nu^2)| \equiv
|\mathcal{K}_{ij}| =\frac{1}{\hbar}\int_0^{\infty} e^{-\frac{\nu^2
t}{\hbar}} K_t(a_i,a_j) \,dt
\leq  C'(\delta) \left[\frac{m
\pi} {\hbar^2}\; K_{0}(\alpha_{ij}\nu) +
\frac{\alpha_{ij}}{4\pi}\; \frac{K_{1}(\alpha_{ij} \nu)}{\nu R^2}
\right]\;,
\end{split}
\ee
where
\be \alpha_{ij}\equiv \sqrt{\frac{8 m\, d_{ij}^2}{D(\delta)
\hbar^2}} \;,\ee
and $K_{0}(x)$, $K_{1}(x)$ are modified Bessel functions. This
shows us that the infinite series in the non-diagonal part of the
characteristic matrix is finite and bounded from above according
to (\ref{phi ij}). In order to find a lower bound for the diagonal
part, denoted by $\mathcal{D}$, of the matrix $\Phi$ for sphere
$\mathbb{S}^2$, we first recall how the diagonal part of the
matrix $\Phi$ appears in (\ref{phi}):
\be \nonumber \mathcal{D}_{i}= \frac{1}{4\pi R^2}\lim_{\Lambda
\rightarrow \infty} \left[ \sum_{l=0}^{\Lambda}\frac{2l+1}
{\hbarmr l(l+1)+\mu_{i}^2} - \sum_{l=0}^{\Lambda}
\frac{2l+1}{\hbarmr l(l+1)+\nu^2} \right] \geq 0 \;. \ee
Instead of calculating explicitly this limit as we have done in section \ref{Delta Interactions on Sphere $S^2$}, we
estimate a lower bound of it by means of integrals replaced by the sums as follows
\be \nonumber \label{intestimates} \mathcal{D}_{i} \geq
\frac{1}{4\pi R^2}\lim_{\Lambda \rightarrow \infty}
\left[\int_{0}^{\Lambda+1}\frac{2t+1}{\hbarmr
t(t+1)+\mu_{i}^2}\;dt - \int_{0}^{\Lambda}\frac{2t+1}{\hbarmr
t(t+1)+\nu^2}\;dt-\frac{1}{\nu^2}\right]\;. \ee
After taking the limit we find
\be \mathcal{D}_{i} \geq \left[\frac{m}{\pi \hbar^2}\log
(\nu/\mu_i)-\frac{1}{4\pi R^2 \nu^2}\right] \;,\ee
and using the estimate for logarithmic functions in
\cite{Abramowitz}
\be \label{logestimates}\log x > \frac{x-1}{x} \hspace{1cm}
\mathrm{for}\; x > 0 \;,x \neq 1 \;, \ee
we obtain
\be \label{intestD} \mathcal{D}_{i} \geq \left[\frac{m}{\pi
\hbar^2}\log (\nu/\mu_i)-\frac{1}{4\pi R^2 \nu^2}\right] >
\left[\frac{m}{\pi \hbar^2}-\frac{m \mu_i }{\pi \hbar^2 \nu}
-\frac{1}{4\pi R^2 \nu^2}\right] > 0 \;. \ee
For positive definiteness, we have assumed $\nu$ is sufficiently
large, which is not a particularly restrictive condition. In fact,
one can try to find sharper estimates by means of the integral
representations of digamma functions (\ref{digammarep1}) and
(\ref{digammarep2}) without this assumption. However, the
estimated functions in this case are too complicated to suggest a
bound for ground state energy.

A well-known theorem in matrix analysis, called Ger\v{s}gorin
Theorem \cite{Gershgorin2} states that all the eigenvalues
$\lambda_i$ of the renormalized matrix $\Phi$ are located in the
union of $N$ discs
\be \label{gerstheorem} \bigcup_{i=1}^{N} \{|\lambda_i-
\Phi_{ii}|\leq R'_{i}(\Phi)\} \equiv G(\Phi) \;, \ee
where $R'_{i}(\Phi)\equiv \sum_{i \neq j =1}^{N} |\Phi_{ij}| $ and
$1\leq i \leq N$. If we want $\lambda=0$ not to be an eigenvalue,
then none of the discs should contain $\lambda=0$. Then, we should
impose
\be \label{gersh} |-\mathcal{D}_{i}(\nu)| > \sum_{i \neq
j}^{N}|\mathcal{K}_{ij}(\nu)| \;, \ee
for all $i$. This is possible for a critical value $\nu
> \nu^{*}$ since the left hand side is an increasing
function of $\nu$ and the right hand side is a decreasing function
of it for a given $d$ and $N$. In fact, this inequality obviously
provide a lower bound for the bound state energy by just plotting
the functions on both sides in spite of how complicated the form
of functions are. However, we shall try to find an explicit
expression for the lower bound of the ground state energy
depending on the number of delta interactions. In order to achieve
this, we choose $\nu$ such that;
\bea \nonumber |-\mathcal{D}_{i}(\nu)| & > & \left[\frac{m}{\pi
\hbar^2}-\frac{m \mu_i }{\pi \hbar^2 \nu} -\frac{1}{4\pi R^2
\nu^2}\right]\\ (N-1) C'(\delta) \left[\frac{m \pi} {\hbar^2}\;
K_{0}(\alpha \nu) + \frac{\alpha}{4\pi}\; \frac{K_{1}(\alpha
\nu)}{\nu R^2} \right] & > & \sum_{i \neq
j}^{N}|\mathcal{K}_{ij}(\nu)|\;, \eea
where we have used the monotonic behavior of the functions in
$\mathcal{D}_{ij}$ and $\mathcal{K}_{ij}$ and defined $\mu \equiv
\max_{i}\mu_i$ and $\alpha \equiv \min_{i\neq j}\alpha_{ij}$ or $d
\equiv \min_{i\neq j} d_{ij}$.
From the integral representations of the Bessel functions for $z
\in \mathbb{R}^{+}$ \cite{Lebedev}
\be \nonumber
\begin{split}
K_{0}(z)&=\int_{0}^{\infty}e^{-z \cosh t}\;dt \;, \\
K_{1}(z)&= z \int_{0}^{\infty} e^{-z \cosh t} \sinh^2 t \; dt \;,
\end{split}
\ee
and using the inequalities $\frac{e^{t}}{2} < \cosh t$, $\sinh^2 t
< \frac{e^{2t}}{4}$ for all $t \in \mathbb{R}^{+}$, we can find
the upper bounds for the functions $K_0$ and $K_1$
\be \label{besselestimates}
\begin{split}
K_{0}(\alpha \nu)& < \frac{2\, e^{-\frac{\alpha \nu}{2}}}{\alpha \nu}\;, \\
K_{1}(\alpha \nu)& < e^{-\frac{\alpha \nu}{2}}
\left(\frac{1}{\alpha \nu}+\frac{1}{2}\right)\;,
\end{split}
\ee
where $\alpha \nu \in \mathbb{R}^{+}$. Considering the estimated
bounds for Bessel functions, it is easy to see that
\be \nonumber
\begin{split}
\frac{m}{\pi \hbar^2}(N-1)C'(\delta)& \left[\frac{2 \pi^2
e^{-\frac{\alpha \nu}{2}}}{\alpha \nu}+ \frac{e^{-\frac{\alpha
\nu}{2}} \mu_{R}^2}{2 \nu^2} + \frac{e^{-\frac{\alpha \nu}{2}}
\mu_{R}^2 \alpha}{4 \nu} \right]
\\ & \qquad \qquad \qquad \qquad \qquad \qquad > (N-1) C'(\delta) \left[\frac{m \pi} {\hbar^2}\; K_{0}(\alpha \nu)
+ \frac{\alpha}{4\pi}\; \frac{K_{1}(\alpha \nu)}{\nu R^2} \right]
\;.
\end{split}
\ee
Using the argument $\nu > \sqrt{2} \mu_R$ in equation
(\ref{intestD}) and last inequality, we impose the following
inequality with the help of Ger\v{s}gorin Theorem:
\be \nu > \mu + \frac{\mu_R}{2 \sqrt{2}} + (N-1)C'(\delta)
e^{-\frac{\alpha \nu}{2}} \left[\frac{2 \pi^2}{\alpha}+
\frac{\mu_{R}}{2 \sqrt{2}} + \frac{\mu_{R}^2 \alpha}{4} \right]
\;. \ee
Let us make the following reasonable assumptions and take these
for granted for the present (we will later show that they indeed
satisfy these conditions by finding the extremum of ground state
energy with respect to the parameter $\delta$)
\bea \frac{\mu_{R}^{2} \alpha}{4} & < & \frac{2 \pi^2}{\alpha}
\;,\\
\frac{\mu_R}{2 \sqrt{2}} & < & \frac{2 \pi^2}{\alpha} \;, \\
\frac{1}{\sqrt{D(\delta)}} & > & \frac{1}{5} \;,\eea
so that the inequality becomes
\be \label{spherein} \nu  > \mu + \frac{\mu_R}{2 \sqrt{2}} + 3
\pi^2 \mu_{d} (N-1)C'(\delta) \sqrt{D(\delta)} \; e^{-
\frac{\nu}{5 \mu_d}} \;,\ee
from which we conclude that there exists a critical value $\nu >
\nu^{*}$ for a given $N$ such that $\lambda \neq 0$ and then, the
ground state energy cannot be less than $- {\nu^*}^{2}$:
\be E_{gr} \geq -{\nu^*}^{2} =- \left\{\mu + \frac{\mu_R}{2
\sqrt{2}} + 5 \mu_{d} W \left[\frac{3 \pi^2}{5}\; C'(\delta)
\sqrt{D(\delta)} (N-1) \; e^{-\frac{\left(\mu + \frac{\mu_R}{2
\sqrt{2}}\right)}{5 \mu_d}} \right]\right\}^{2}\;, \ee
where $W$ is the Lambert W-function, also called Omega function or
product-log function \cite{Corless}. Now, we choose $\delta$ in
such a way that the energy bound take its minimum value. This is
accomplished if $\delta$ is chosen approximately $0.508$, which is
independent of the parameters in the problem. This independence
can be easily realized from the form of inequality
(\ref{spherein}). By substituting the values of $C'(\delta)$ and $
D(\delta)$, we estimate a lower bound for the ground state energy:
\be \label{Egrsphere} E_{gr} \geq -{\nu^*}^{2} = - \left\{\mu +
\frac{\mu_R}{2 \sqrt{2}} + 5 \mu_{d} W \left[28 \pi^2 \; (N-1) \;
e^{-\frac{\left(\mu + \frac{\mu_R}{2 \sqrt{2}}\right)}{5 \mu_d}}
\right]\right\}^{2} \;.\ee
By using this value of $\delta$ and the fact that $d < 2 \pi R$,
the consistency of the assumption we made can be shown easily.
Finally, we shall consider the large $N$ behavior of the ground
state energy. The asymptotic expansion of product-log function W
\cite{Corless} for large $z$ is given as
\be \label{prodlogas} W(z) \sim \log z - \log\log z\;. \ee
Hence, this leads to
\be E_{gr} \sim -  \mu_d^2  \left[\log \left( N \right) - \log\log
\left( N \right)\right]^{2} \;.\ee

The method we have introduced for the two dimensional sphere $\mathbb{S}^{2}$ can also be applied to a general compact
manifold. The main idea is based on finding an upper and lower bound for  the characteristic matrix or heat kernel
(based on the work by Li and Yau). Then, Ger\v{s}gorin theorem allows us to estimate a lower bound for the ground state
energy.

\section{\label{delta on hyperbolic} Finitely Many Dirac-Delta Interactions on Hyperbolic Spaces}

The hyperbolic space $\mathbb{H}^{n}$ is defined as maximally
symmetric and simply connected complete $n$-dimensional Riemannian
manifold with a constant negative sectional curvature $-1/R$,
which is also in some sense considered to be the negative
curvature analog of the sphere $\mathbb{S}^n$. We shall deal with
the delta interactions on the hyperbolic spaces $\mathbb{H}^3$ and
$\mathbb{H}^2$ in the following sections. The method developed in
the previous sections for $\mathbb{S}^2$ will be useful as well
for the hyperbolic spaces. The heat kernel on hyperbolic spaces
\cite{heatkernelhyperbolic}, written in terms of dimensionless
quantities:
\be
\begin{split} K_{t}(x,y) & = \frac{1}{(4\pi t)^{3/2}} \frac{d}{\sinh d}e^{-t-\frac{d^2}{4 t}} \hspace{1cm}\mathrm{on}
\;\;\;\mathbb{H}^3\;,
\\K_{t}(x,y) & = \frac{\sqrt{2}}{(4\pi t)^{3/2}}e^{-\frac{t}{4}}\int_{d}^{\infty}\frac{s \; e^{-\frac{s^2}{4t}}}
{(\cosh s- \cosh d)^{1/2}}\;d s \hspace{1cm}\mathrm{on}
\;\;\;\mathbb{H}^2\;,
\end{split}
\ee
where $d \equiv \mathrm{dist}(x,y)$, geodesic distance between two
points $x$ and $y$ on $\mathbb{H}^n$.

Although spectral theorem and asymptotic expansion of heat kernel discussed in the previous sections may not be valid
for general non-compact manifolds, we shall demonstrate that for the specific examples in non-compact manifolds, such
as $\mathbb{H}^2$ and $\mathbb{H}^3$, our viewpoint still works. It would be desirable to show the equivalence between
the eigenfunction expansion and the heat kernel method for the regularization in non-compact manifolds rigourously.
Nevertheless, we have not been able to do this. The main idea is similar in spirit to the renormalization procedure
introduced for the compact manifolds.

\subsection{\label{delta on H^3} Finitely Many Dirac-Delta Interactions on Hyperbolic
Space $\mathbb{H}^{3}$}

In the hyperbolic space $\mathbb{H}^3 = \{x \in \mathbb{R}^3 | x_3
> 0 \}$, the geodesic distance $d$ is defined as
\be \nonumber \cosh \frac{d(x,y)}{R} = 1 + \frac{|x-y|^2}{2 \; x_3
\; y_3}, \ee
where $R$ is the scaling parameter. The Schr\"{o}dinger equation
for the bound states of a particle living on $\mathbb{H}^3$ under
the influence of $N$ attractive delta interactions is
\be \left[\hbarm \triangle_{\mathbb{H}^3} - \sum_{i=1}^N \; g_i \;
\delta^3(\chi-\chi_i, \theta-\theta_i, \phi-\phi_i)  \right] \psi
= -\nu^2 \psi, \ee
where Laplacian $\triangle_{\mathbb{H}^3}$ in polar coordinates
$(\chi,\theta,\phi)$
\be \triangle_{\mathbb{H}^3}
=-\frac{1}{R^3}\frac{\partial^2}{\partial \psi^2}- \frac{2\coth
\psi}{R^3}
 \frac{\partial}{\partial \psi}+\frac{1}{R \sinh^2 \psi}
\Delta_{\mathbb{S}^2} \;. \ee
We have an explicit formula \cite{heatkernelhyperbolic} for the
heat kernel of the three dimensional hyperbolic plane
$\mathbb{H}^3$ written by using physical constants
\be K_t(x, y) = \frac{1}{R^3}\frac{\frac{d(x,y)}{R}}{\left(4 \pi
\left[\frac{\hbar}{2mR^2}\right] t\right)^{3/2} \sinh
\frac{d(x,y)}{R}} \exp \left(-\frac{\hbar t}{2mR^2}  - \frac{m
d(x,y)^2}{2 \hbar t}\right) \ee
such that as $R \rightarrow \infty$, we can obtain the heat kernel
on $\mathbb{R}^3$. Hence we have the free resolvent kernel as
\be \label{H^3}
\begin{split}
\langle a_i | \left( H_0 - z \right)^{-1}|a_j \rangle & =
\frac{1}{\hbar R^3} \int_0^{\infty} \frac{\frac{d_{ij}}{R} \; \exp
\left( \frac{z t}{\hbar} -\frac{\hbar t}{2mR^2}  - \frac{m
d_{ij}^2}{2 \hbar t}\right)}{\left(4 \pi
\left[\frac{\hbar}{2mR^2}\right]
t\right)^{3/2} \sinh \frac{d_{ij}}{R}} \, dt \\
&= \left[\frac{1}{4\pi R^3} \; \frac{\frac{d_{ij}}{R}}{\sinh
\frac{d_{ij}}{R}} \exp \left(-\frac{\mu_R}{\mu_{d_{ij}}} \sqrt{1-
\frac{z}{\mu_{R}^2}}\right) \right] \;
\frac{\mu_{d_{ij}}}{\mu_R^3} ,
\end{split}
\ee
where $d_{ij} \equiv d(a_i,a_j)$, $\mu_R^2 \equiv \frac{\hbar^2}{2 m R^2}$, $\mu_{d_{ij}}^2 \equiv \frac{\hbar^2}{2 m
d_{ij}^2}$. It follows easily that this term gives infinity when $i=j$, that is, the diagonal term in the
characteristic matrix is divergent. Then, we can now proceed the regularization and renormalization schemes analogously
for the hyperbolic spaces. However, the divergence in hyperbolic space $\mathbb{H}^3$ is due to fact that the lower
bound $t$ of integral (\ref{H^3}) is zero. Hence we regularize the divergent term by introducing a lower cut-off
$\epsilon$, as we have shown in section \ref{phi and heat kernel}, we expect this should in some way related to the
ultraviolet regularization. We next define the coupling constant as a function of this cut-off:
\be \nonumber \Phi_{ii}(z) = \lim_{\epsilon \rightarrow 0^+}
\left[ g_i^{-1}(\epsilon) - \frac{1}{(4\pi)^{3/2} \mu_R^2 R^3} \;
\int_\epsilon^\infty u^{-3/2}
e^{-\left[1-\frac{z}{\mu_R^2}\right]u }\;du \right]\;, \ee
where the integration variable $u \equiv \hbarmr \, t$ is
introduced for simplicity. The natural choice for
$g_i^{-1}(\epsilon)$ is simply
\be \nonumber g_i^{-1}(\epsilon) = \frac{1}{(4\pi)^{3/2} \mu_R^2
R^3} \; \int_\epsilon^\infty u^{-3/2}
e^{-\left[1+\frac{\mu_i^2}{\mu_R^2} \right]u}\;du, \ee
where $\mu_i$ is an experimentally measured bound state energy for the single delta interaction and it helps us to keep
track of the strength of point interactions. In $\epsilon \rightarrow 0^+$ limit, we have found the explicit
renormalized characteristic matrix for $\mathbb{H}^3$
\be \Phi_{ij}(z) = \frac{1}{4 \pi} \, \frac{1}{\mu_R^2 R^3}
\begin{cases}
\begin{split}
\sqrt{1-\frac{z\phantom{^2}}{\mu_R^2}}-\sqrt{1+\frac{\mu_i^2}{\mu_R^2}}
\end{split}
& \quad \textrm{if $i=j$} \\ \\
\begin{split}
- \frac{\mu_{d_{ij}}}{\mu_R} \, \frac{\frac{d_{ij}}{R}}{\sinh
\frac{d_{ij}}{R}} \; \exp \left(-\frac{\mu_R}{\mu_{d_{ij}}}
\sqrt{1- \frac{z}{\mu_{R}^2}}\right)
\end{split}
& \quad \textrm{if $i \neq j$}.
\end{cases}
\ee
Then, we have the resolvent equation (\ref{resolvent}) with the free resolvent kernel $R_0(x,y|z)$ for $\mathbb{H}^3$
given by
\be R_0(x,y|z) = \frac{1}{4 \pi} \, \frac{1}{\mu_R^2 R^3}
\frac{\mu_{d(x,y)}}{\mu_R} \, \frac{\frac{d(x,y)}{R}}{\sinh
\frac{d(x,y)}{R}} \; \exp \left(-\frac{\mu_R}{\mu_{d(x,y)}}
\sqrt{1- \frac{z}{\mu_{R}^2}}\right)\;, \ee
from which we can get all information about the system. Using the Ger\v{s}gorin Theorem (\ref{gerstheorem}) for this
matrix, and following the same ideas introduced for $\mathbb{S}^2$ we obtain
\be \nonumber
\left[\sqrt{1+\frac{\nu^2}{\mu_R^2}}-\sqrt{1+\frac{\mu^2}{\mu_R^2}}\right]
> (N-1) \frac{\mu_{d}}{\mu_R}\frac{d}{R \sinh \frac{d}{R}}
\exp \left(-\frac{\mu_R}{\mu_{d}} \sqrt{1+
\frac{\nu^2}{\mu_{R}^2}}\right)\;.
\ee
where we have taken $z=-\nu^2$ and chosen $d \equiv \min_{i\neq j} d_{ij}$, and $\mu \equiv \max_i \mu_i$. It turns out
that this inequality indicates that there exist a critical $\nu > \nu^* $ for a given $d$ and $N$ for which this
inequality is satisfied and zero is not an eigenvalue. Therefore, the ground state energy cannot be less than $-
{\nu^*}^2$:
\be
\begin{split}
E_{gr} \geq - {\nu^*}^2 =-\mu^2 - 2 \mu_d \sqrt{\mu_{R}^2+\mu^2}\;
& W \left[\frac{e^{-\frac{\mu_R}{\mu_{d}} \sqrt{1+
\frac{\mu^2}{\mu_{R}^2}}}\;d(N-1)}{R \sinh \frac{d}{R}}\right] \\
& \qquad \qquad \qquad - \mu_{d}^2 \; W
\left[\frac{e^{-\frac{\mu_R}{\mu_{d}} \sqrt{1+
\frac{\mu^2}{\mu_{R}^2}}}\;d(N-1)}{R \sinh \frac{d}{R}}\right]^{2}
\;.
\end{split}
\ee
For the large $N$ behavior of the ground state energy, the
estimate becomes
\bea \nonumber E_{gr}  \sim & - 2 \mu_d
\sqrt{\mu_{R}^2+\mu^2}\left[\log N- \log\log N\right] - \mu_{d}^2
\left[ \log N - \log\log N \right]^2\;. \eea
Now let us consider the two center case on the hyperbolic plane
$\mathbb{H}^3$ and assume again that their strengths (or bound
state energies of each center) are the same. In this way,
determining equation ($\det \Phi =0$) becomes
\be \nonumber
\sqrt{1+\frac{\nu^2}{\mu_R^2}}-\sqrt{1+\frac{\mu^2}{\mu_R^2}} =
\pm \frac{\mu_{d}}{\mu_R} \, \frac{\frac{d}{R}}{\sinh \frac{d}{R}}
\; \exp \left(-\frac{\mu_R}{\mu_{d}} \sqrt{1+
\frac{\nu^2}{\mu_{R}^2}}\right). \ee
If we expand it for small $d$ we have
\be \nonumber
\sqrt{1+\frac{\nu^2}{\mu_R^2}}-\sqrt{1+\frac{\mu^2}{\mu_R^2}} =
\pm \frac{\mu_{d}}{\mu_R} \left[1 -
\sqrt{\frac{\mu_R^2}{\mu_{d}^2}+ \frac{\nu^2}{\mu_{d}^2}} \right],
\ee
from which we can conclude
\be \nonumber E_{gr} = -\nu^2 \simeq \frac{3}{4}\, \mu_R^2 -
\frac{\mu^2}{4} - \frac{\mu_{d}^2}{4} - \frac{\mu_{d} \, \mu_R}{2}
\sqrt{1+\frac{\mu^2}{\mu_R^2}}\;. \ee
Similarly, for large values of $d$, the right hand side of the
energy equation for two dirac delta interactions vanishes, so that
we obtain the ground state energy $E_{gr}=-\nu^2 = -\mu^2$.

\subsection{\label{delta on H^2} Finitely Many Dirac-Delta Interactions on Hyperbolic
Plane $\mathbb{H}^2$}

The geodesic distance on the hyperbolic plane $\mathbb{H}^2$ is
defined by
\be \nonumber \cosh \frac{d(x,y)}{R} = 1 + \frac{|x-y|^2}{2 \; x_2
\; y_2}\;, \ee
where $R$ is a scale distance. Then, the Schr\"{o}dinger equation
for the bound states of a particle living on $\mathbb{H}^2$ in the
presence of $N$ attractive delta interactions is
\be \left[\hbarm \triangle_{\mathbb{H}^2} - \sum_{i=1}^N \; g_i \;
\delta^2(\theta-\theta_i, \phi-\phi_i)  \right] \psi = -\nu^2
\psi, \ee
where the Laplacian $\triangle_{\mathbb{H}^2}$ in polar
coordinates $(\theta,\phi)$ is given by
\be \triangle_{\mathbb{H}^2} =
-\frac{1}{R^2}\frac{\partial^2}{\partial \theta^2}- \frac{2\coth
\theta}{R^2}
 \frac{\partial}{\partial \theta}-\frac{1}{R^2 \sinh^2 \theta}
\frac{\partial^2}{\partial\phi^2}\;. \ee
The heat kernel for $\mathbb{H}^2$ \cite{heatkernelhyperbolic}
with the proper physical parameters is
\be K_t(x,y) = \frac{\sqrt{2}}{(4 \pi
\left[\frac{\hbar}{2mR^2}\right] t)^{3/2}} \;
\frac{e^{-\frac{\hbar}{2mR^2} \, \frac{t}{4}}}{R^2} \,
\int_{\frac{d(x,y)}{R}}^\infty \frac{r e^{-\frac{r^2}{4} \frac{2m
R^2}{\hbar} \frac{1}{t}}}{\sqrt{\cosh r - \cosh \frac{d(x,y)}{R}}}
\; dr \;. \ee
One can check that this goes to the heat kernel on $\mathbb{R}^2$
as $R\rightarrow\infty$. Then, the free resolvent kernel is
immediately obtained
\be
\begin{split}
\langle a_i | \left( H_0 - z \right)^{-1} & |a_j \rangle  =
\frac{1}{\hbar R^2} \int_0^{\infty} e^{z \frac{t}{\hbar}}
\frac{\sqrt{2}}{(4 \pi \left[\frac{\hbar}{2mR^2}\right] t)^{3/2}}
\; e^{-\frac{\hbar}{2mR^2}\, \frac{t}{4}}\;
\left[\int_{\frac{d_{ij}}{R}}^\infty \frac{r e^{-\frac{r^2}{4}
\frac{2m R^2}{\hbar} \frac{1}{t}}}{\sqrt{\cosh r - \cosh
\frac{d_{ij}}{R}}} \; dr \right] \, dt \nonumber
\\
& = \frac{1}{4\pi \, \mu_R^2 R^2} \;
\int_{\frac{d_{ij}}{R}}^\infty \frac{e^{-\frac{1}{2} r
\sqrt{1-\frac{4z}{\mu_R^2}}}}{\sqrt{\cosh r - \cosh
\frac{d_{ij}}{R}}} \; dr.
\end{split}
\ee
We see that the diagonal term, which corresponds to $d_{ij}=0$ is
divergent, as expected. Therefore we again repeat the similar
regularization and renormalization procedure as we have done for
$\mathbb{H}^3$. After introducing a cut-off to the lower limit of
the integral
\be \nonumber
\begin{split}
\Phi_{ii}(z) &= \lim_{\epsilon \rightarrow 0^+} \left[
g_i^{-1}(\epsilon) - \frac{1}{4\pi \, \mu_R^2 R^2} \;
\int_{\epsilon}^\infty \frac{e^{-\frac{1}{2} r
\sqrt{1-\frac{4z}{\mu_R^2}}}}{\sqrt{\cosh r - 1}} \; dr \right] \\
& = \lim_{\epsilon \rightarrow 0^+} \left[ g_i^{-1}(\epsilon) -
\frac{\sqrt{2}}{4\pi \, \mu_R^2 R^2} \;
\int_{\frac{\epsilon}{2}}^\infty \frac{e^{- u
\sqrt{1-\frac{4z}{\mu_R^2}}}}{\sinh u} \; du \right].
\end{split}
\ee
and by the natural choice for $g_i^{-1}(\epsilon)$
\be \nonumber g_i^{-1}(\epsilon) = \frac{\sqrt{2}}{4\pi \, \mu_R^2
R^2} \; \int_{\frac{\epsilon}{2}}^\infty \frac{e^{- u
\sqrt{1+\frac{4 \mu_i^2}{\mu_R^2}}}}{\sinh u} \; du \;, \ee
we have obtained the renormalized characteristic matrix for
$\mathbb{H}^2$ in the $\epsilon \rightarrow 0$ limit,
\be \label{phih^2} \Phi_{ij}(z) = \frac{1}{4 \pi R^2} \,
\frac{1}{\mu_R^2}
\begin{cases}
\begin{split}
\sqrt{2} \left[ \psi\left(\frac{1}{2} +  \sqrt{\frac{1}{4}-
\frac{z}{\mu_R^2}}\right) - \psi\left(\frac{1}{2} +
\sqrt{\frac{1}{4}+ \frac{\mu_i^2}{\mu_R^2}}\right) \right]
\end{split}
& \quad \textrm{if $i=j$} \\ \\
\begin{split}
- \int_{\frac{d_{ij}}{R}}^\infty \frac{e^{-\frac{1}{2} r
\sqrt{1-\frac{4z}{\mu_R^2}}}}{\sqrt{\cosh r - \cosh
\frac{d_{ij}}{R}}} \; dr
\end{split}
& \quad \textrm{if $i \neq j$},
\end{cases}
\ee
where $\psi$ is the digamma function. Then, we have the resolvent
equation (\ref{resolvent}) in which the free resolvent kernel
$R_0(x,y|z)$ for $\mathbb{H}^2$ is given by
\be R_0(x,y|z) = \frac{1}{4 \pi R^2} \, \frac{1}{\mu_R^2}
\int_{\frac{d(x,y)}{R}}^\infty \frac{e^{-\frac{1}{2} r
\sqrt{1-\frac{4z}{\mu_R^2}}}}{\sqrt{\cosh r - \cosh
\frac{d(x,y)}{R}}} \; dr. \ee
The integral on the right hand side is in fact one of the integral
representation of the Legendre polynomials of second type
\cite{Lebedev}
\be \sqrt{2} Q_{\lambda}\left(\cosh \frac{d(x,y)}{R}\right) =
\int_{\frac{d(x,y)}{R}} ^\infty \frac{e^{-(\lambda +
\frac{1}{2})r}}{{\sqrt{\cosh r - \cosh \frac{d(x,y)}{R}}}} \; dr
\;,\ee
which are defined for $\Re \left(\lambda \right) > -1$ and in our
case $ \Re \left(\lambda \right) = \Re
\left(\frac{1}{2}\sqrt{1-\frac{4 z}{\mu_{R^2}}}-\frac{1}{2}\right)
> -1$. Therefore, the free resolvent in terms of $Q_{\lambda}$
\be R_0(x,y|z) = \frac{1}{4 \pi R^2} \, \frac{1}{\mu_R^2} \sqrt{2}
Q_{\frac{1}{2}\sqrt{1-\frac{4 z}{\mu_{R^2}}}-\frac{1}{2}
}\left(\cosh \frac{d(x,y)}{R}\right) \;.\ee
Ger\v{s}gorin theorem allows us to estimate the lower bound for
the bound state energy as done for $\mathbb{S}^2$ and
$\mathbb{H}^3$. In order not to have zero as an eigenvalue, we
must have
\be \label{w} \sqrt{2} \left[ \psi\left(\frac{1}{2} +
\sqrt{\frac{1}{4}+ \frac{\nu^2}{\mu_R^2}}\right) -
\psi\left(\frac{1}{2} + \sqrt{\frac{1}{4}+
\frac{\mu_i^2}{\mu_R^2}}\right) \right] > \sum_{i\neq j}
\int_{\frac{d_{ij}}{R}}^\infty \frac{e^{-\frac{1}{2} r
\sqrt{1+\frac{4\nu^2}{\mu_R^2}}}}{\sqrt{\cosh r - \cosh
\frac{d_{ij}}{R}}} \; dr \;, \ee
for all $i$ and we have taken $z=-\nu^2$ and $\nu > \max_i \mu_i
$. It is easy to see this inequality is satisfied for some values
of $\nu$ because the left hand side is an increasing function,
whereas the right hand side is a decreasing function of $\nu$.
However, it is not so easy to give an explicit estimate for $\nu$
by this inequality so we will estimate the functions on both
sides. The inequality for the digamma functions \cite{Alzer}
\be \psi(x) > \log x - \frac{1}{x} \hspace{1cm}x>0 \;, \ee
which can be obtained from the integral representation
(\ref{digammarep2}), and $\frac{1}{2} + \sqrt{\frac{1}{4}+ x^2}
\geq x$ for all $x>0$ helps us that we can find the following
inequality by assuming $\nu$ is sufficiently large
\be \label{h2}
\left[ \psi\left(\frac{1}{2} + \sqrt{\frac{1}{4}+
\frac{\nu^2}{\mu_R^2}}\right) - \psi\left(\frac{1}{2} +
\sqrt{\frac{1}{4}+ \frac{\mu_i^2}{\mu_R^2}}\right) \right] >
\left[\log \frac{\nu}{\mu_R}- \frac{\mu_R}{\nu} -
\psi\left(\frac{1}{2} + \sqrt{\frac{1}{4}+
\frac{\mu_{i}^2}{\mu_R^2}}\right) \right]  \;. \ee
Since the right hand side of equation (\ref{w}) is $(N-1) \sqrt{2}
\, Q_{\lambda}\left(\cosh \frac{d_{ij}}{R}\right)$ we can find an
upper bound for this function, using another integral
representation of the second type Legendre polynomials
\cite{Lebedev}:
\be Q_{\lambda}\left(\cosh \frac{d_{ij}}{R}\right) =
\frac{1}{\Gamma(\lambda+1)} \int_{0}^{\infty} e^{-t \cosh
\frac{d_{ij}}{R}} K_{0}\left(t \sinh \frac{d_{ij}}{R}\right)
t^{\lambda} \;dt \;,\ee
where $\Im \left(\frac{d_{ij}}{R}\right) = 0$ and $\lambda =
\frac{1}{2}\sqrt{1+ \frac{4 \nu^2}{\mu_R^{2}}}- \frac{1}{2}$.
Using the estimate for the function $K_0$ given in equation
(\ref{besselestimates}), we obtain
\be \nonumber \sqrt{2} Q_{\lambda}\left(\cosh
\frac{d_{ij}}{R}\right) < \frac{2 \sqrt{2}}{\Gamma(\lambda+1)
\sinh \frac{d_{ij}}{R}} \int_{0}^{\infty} e^{-t \left(\cosh
\frac{d_{ij}}{R} + \frac{1}{2} \sinh \frac{d_{ij}}{R} \right)}
t^{\lambda-1} \;dt \ee
and the right hand side is just the Gamma function, then the
estimate becomes
\be \nonumber \sqrt{2} Q_{\lambda}\left(\cosh
\frac{d_{ij}}{R}\right) < \frac{2
\sqrt{2}\Gamma(\lambda)}{\Gamma(\lambda+1) \left(\cosh
\frac{d_{ij}}{R} + \frac{1}{2} \sinh \frac{d_{ij}}{R}
\right)^{\lambda} \sinh \frac{d_{ij}}{R}} \;.\ee
Using identity $\Gamma(\lambda+1)= \lambda \Gamma(\lambda)$ and
the assumption $\nu/\mu_{R}>1 $ and
$\sqrt{1+\frac{4\nu^2}{\mu_{R}^{2}}} > \frac{2\nu}{\mu_R}$, we get
\be
\begin{split}
\sqrt{2} Q_{\lambda}\left(\cosh \frac{d_{ij}}{R}\right) & <
\frac{4 \sqrt{2}}{(\frac{2\nu}{\mu_R}-1)} \frac{1}{\left(\cosh
\frac{d_{ij}}{R} + \frac{1}{2} \sinh \frac{d_{ij}}{R}
\right)^{\lambda} \sinh \frac{d_{ij}}{R}} \\ & < \frac{4
\sqrt{2}}{\frac{\nu}{\mu_R}}
\frac{e^{-(\frac{\nu}{\mu_R}-\frac{1}{2}) \log \left(\cosh
\frac{d_{ij}}{R}+ \frac{1}{2} \sinh \frac{d_{ij}}{R} \right)
}}{\sinh \frac{d_{ij}}{R}} \;.
\end{split}
\ee
Also, by choosing $d \equiv \min_{i\neq j} d_{ij}$ and $\mu \equiv
\max_i \mu_i$, we easily find
\be
 \left[\log \frac{\nu}{\mu_R}-
\frac{\mu_R}{\nu} - \psi\left(\frac{1}{2} + \sqrt{\frac{1}{4}+
\frac{\mu_i^2}{\mu_R^2}}\right) \right] >  \left[\log
\frac{\nu}{\mu_R}- \frac{\mu_R}{\nu} - \psi\left(\frac{1}{2} +
\sqrt{\frac{1}{4}+ \frac{\mu^2}{\mu_R^2}}\right) \right] \;, \ee
and
\be \frac{e^{-(\frac{\nu}{\mu_R}-\frac{1}{2}) \log \left(\cosh
\frac{d_{ij}}{R}+ \frac{1}{2} \sinh \frac{d_{ij}}{R} \right)
}}{\sinh \frac{d_{ij}}{R}}  <
\frac{e^{-(\frac{\nu}{\mu_R}-\frac{1}{2}) \log \left(\cosh
\frac{d}{R}+ \frac{1}{2} \sinh \frac{d}{R} \right) }}{\sinh
\frac{d}{R}}  \;. \ee
Therefore, we impose
\be \left[\log \frac{\nu}{\mu_R}- \frac{\mu_R}{\nu} -
\psi\left(\frac{1}{2} + \sqrt{\frac{1}{4}+
\frac{\mu^2}{\mu_R^2}}\right) \right] >  \frac{4(N-1)
}{\frac{\nu}{\mu_R}} \frac{e^{-(\frac{\nu}{\mu_R}-\frac{1}{2})
\log \left(\cosh \frac{d}{R}+ \frac{1}{2} \sinh \frac{d}{R}
\right) }}{\sinh \frac{d}{R}} \;. \ee
It is immediately seen that there exists a critical value $\nu >
\nu^*$ for a given $d$ and $N$ for which this inequality is
satisfied and zero is not an eigenvalue. Last inequality can be
written as
\be \nonumber e^{(\frac{\nu}{\mu_R}-\frac{1}{2}) \log \left(\cosh
\frac{d}{R}+ \frac{1}{2} \sinh \frac{d}{R} \right) }
\left(\frac{\nu}{\mu_R} \log \left(\frac{\nu}{\mu_R \;
e^{\psi\left(\frac{1}{2} + \sqrt{\frac{1}{4}+
\frac{\mu^2}{\mu_R^2}}\right)}} \right) -1 \right)> \frac{4
(N-1)}{\sinh \frac{d}{R}}\,. \ee
If $\frac{\nu}{\mu_R} > e^{\psi\left(\frac{1}{2} +
\sqrt{\frac{1}{4}+ \frac{\mu^2}{\mu_R^2}}\right)+1}$ (independent
of $N$), then we have the lower bound of the ground state energy
\be E_{gr}  \geq - {\nu^*}^2 = - \mu_R^2 \Biggr[ \frac{A +
W\left(\frac{4 (N-1) A e^{-A/2}}{\sinh \frac{d}{R}} \right)}{A}
\Biggl]^{2} \;,\ee
where we define $A \equiv \log \left(\cosh \frac{d}{R}+
\frac{1}{2} \sinh \frac{d}{R} \right) $ for simplicity of
notation. For large values of $N$ as long as the ratio
$\frac{\mu}{\mu_R}$ and $\frac{d}{R}$ is finite, the behavior of
the bound state energy is given by
\be E_{gr} \sim - \mu_R^2 \left[ \frac{\log N -\log\log N}{\log
\left(\cosh \frac{d}{R}+  \frac{1}{2} \sinh \frac{d}{R} \right)}
\right]^2 \;. \ee
This problem again is an example of a certain kind of dimensional
transmutation in non-relativistic quantum mechanics. By
dimensional analysis, the hamiltonian of the system contains
intrinsic energy scales $\frac{\hbar^2}{m d_{ij}^2}$ and
$\hbarmr$. However, after the renormalization, we obtain new
parameters $\mu_i^2$ with energy dimensions. Hence, the number of
parameters we expect for the energy at the beginning has changed
after the renormalization. As it happens  in the $\mathbb{S}^2$
case, the delta potentials on $\mathbb{H}^2$ is an example of a
generalized dimensional transmutation.

\section{\label{conclusion} Conclusion}

In this work, we studied a particle moving  under the influence of  $N$ attractive Dirac delta interactions on some special Riemannian
manifolds. We renormalized the problem and find a finite dimensional matrix $\Phi$, called the characteristic matrix,
by means of which a  well defined expression for the resolvent can be written. All the information about the bound
states can be obtained from the characteristic matrix. The renormalization  can be done  by means of the heat kernel and this is equivalent to the sharp cut-off method for the eigenvalues of the Laplacian, in the case of compact manifolds. We have studied the problem on particular compact and
non-compact manifolds, $\mathbb{S}^2$, $\mathbb{H}^2$, and $\mathbb{H}^3$ and we give explicit lower bound estimates on the
bound state energies for each problem. Although we are concerned with  particular manifolds,
the  basic idea for the renormalization  can be applied also to general
manifolds.

\section{Acknowledgments}

This work is largely based on parts of the master thesis of Bar\i
\c{s} Altunkaynak. The authors gratefully acknowledges the many
helpful discussions of \"{O}. F. Day\i, E. Demiralp, I. H. Duru,
M. Mungan, and H. Uncu. O. T. T. 's research is partially
supported by the Turkish Academy of Sciences, in the frame work of
Young Scientist Award Program (OTT-TUBA-GEBIP/2002-1-18).


\begin{thebibliography}{99}

\bibitem{BFT}H. Bethe and R. Peierls,
\textit{Proceedings of Royal Society}, (London, 1935) Vol 148A,
pp. 146-156; E. Fermi, Ricerca Scientifica \textbf{7}, 13 (1936);
L. H. Thomas, Physical Review, \textbf{47}, 903 (1935).

\bibitem{Berezin} F. A. Berezin and L. D. Faddeev, Soviet Math. Dokl.
\textbf{2}, 372 (1961).

\bibitem{Thorn} C. Thorn, Physical Review D,
\textbf{19}, 639 (1979)

\bibitem{Jackiw} R. Jackiw, \textit{Delta-Function Potentials in Two- and Three-Dimensional
Quantum Mechanics}, (M. A. B. B\'{e}g Memorial Volume, World
Scientific, Singapore, 1991).

\bibitem{Perez} J. Fernando Perez, F. A. B. Coutinho, Am. J. Phys., \textbf{59}, 52 (1991).

\bibitem{Tarrach1} P. Gosdzinsky, R. Tarrach, Am. J. Phys.,  \textbf{59}, 70 (1991).

\bibitem{Mead} L. R. Mead, J. Godines, Am. J. Phys., \textbf{59}, 935 (1991).

\bibitem{Tarrach2} C. Manuel, R. Tarrach, Phys. Lett. B., \textbf{328}, 113 (1994).

\bibitem{Adhikari} S. K. Adhikari, T. Frederico, Phys. Rev. Lett., \textbf{74}, 4572 (1995).

\bibitem{Park} D. K. Park, J. Math. Phys., \textbf{36}, 5453 (1995).

\bibitem{Mitra} I. Mitra, A. DasGupta, B. Dutta-Roy, Am. J. Phys. \textbf{66}, 12 (1998).

\bibitem{Albeverio 2004} S. Albeverio, et al., \textit{Solvable
Models in Quantum Mechanics} (AMS, Providence, Rhode Island,
2004).

\bibitem{Nyeo} Su-Long Nyeo, Am. J. Phys., \textbf{68}, 6 (2000).

\bibitem{Coleman} S. Coleman, E. Weinberg, Phys. Rev. D.
\textbf{7} 1888 (1973).


\bibitem{Huang} K. Hunag, \textit{Quarks, Leptons and Gauge
fields}, (World Scientific, Singapore, 1982).

\bibitem{Camblong} H. E. Camblong, L. N. Epele, H. Fanchiotti, C.
A. G. Canal, Annals Phys. \textbf{287}, 14 (2001); Annals Phys.
\textbf{287}, 57 (2001).


\bibitem{Rajeev2} S. G. Rajeev, Mittag - Leffler Institute preprint ML-7-99, Feb 1999, 63pp (e-Print Archieve: hep-th/9902025).

\bibitem{de witt} B. S. DeWitt, Rev. Mod. Phys., \textbf{29}, 377 (1957);
C. DeWitt-Morette, K. D. Elworthy, B. L. Nelson, and G. S.
Sammelman, Ann. Inst. Henri Poincar\'{e} XXXII, \textbf{4}, 327
(1980); B. S. DeWitt, \textit{Supermanifolds}, (2nd Edition,
Cambridge University Press, Cambridge, 1992); H. Kleinert, Phys.
Lett. B., \textbf{236}, 315 (1990).

\bibitem{Rosenberg} S. Rosenberg, \textit{The Laplacian on Riemannian
Manifold}, (Cambridge University Press, 1998).

\bibitem{Davies} E. B. Davies, \textit{Heat kernels and spectral theory}, (Cambridge University Press, Cambridge
- NewYork - New Rochelle 1989)

\bibitem{Weyl} H. Weyl, Nachr. Akad. Wiss Gottingen Math.- Phys. KL, II, 110, (1911).

\bibitem{Albeverio 2000} S. Albeverio, et al., \textit{Singular Perturbations of Differential Operators
Solvable Schrvdinger-type Operators}, (Cambridge University Press,
Cambridge, 2000).

\bibitem{specialfunc} W. Magnus and F. Oberhettinger,  \textit{Formulas and Theorems for the Special Functions
of Mathematical Physics}, (Springer-Verlag, New York, 3rd ed
1966).

\bibitem{liyau par} Li Peter and Shing-Tung Yau, Acta Mathematica, \textbf{156}, 153 (1986).

\bibitem{Abramowitz} M. Abramowitz, I. A. Stegun, \textit{Handbook of Mathematical Functions with Formulas, Graphs and
Mathematical Tables}, (Tenth Printing ,1972).

\bibitem{Gershgorin2} A. Horn Roger, R. Johnson Charles, \textit{Matrix Analysis},
(Cambridge University Press, 1992).

\bibitem{Lebedev} N. N. Lebedev, \textit{Special Functions and Their
Applications}, (Printice Hall, 1965).

\bibitem{Corless} R. M. Corless, G. H. Gonnet, D. E. G. Hare, D. J. Jeffrey, and D. E. Knuth
Advances in Computational Mathematics \textbf{5}, 329 (1996).

\bibitem{heatkernelhyperbolic} A. Debiard, B. Gaveau, E. Mazet, Th\`{e}or\'{e}mes de
comparison in g\`{e}om\`{e}trie riemannienne, Publ. Kyoto Univ.,
\textbf{12}, 391 (1976); H. P. McKean, J. Diff. Geom., \textbf{4},
359 (1970); E. B. Davies, N. Mandouvalos, Proc. London Math. Soc.,
\textbf{57}, 182 (1988).

\bibitem{Alzer} H. Alzer, Mathematics of Computation, \textbf{66}, 217 (1997).

\end{thebibliography}
\end{document}